\newcommand{\erg}{\mbox{\ensuremath{\mathrm{~erg}}}}
\newcommand{\ergps}{\ensuremath{\mathrm{\erg \ps}}}
\newcommand{\lha}{\ensuremath{L_{\halpha}}}
\newcommand{\xspec}{\textsc{XSPEC}}
\newcommand{\Caldb}{\textsc{CALDB}}
\newcommand{\Ciao}{\textsc{CIAO}}
\newcommand{\km}{\mbox{\ensuremath{\mathrm{~km}}}}
\newcommand{\ps}{\ensuremath{\mathrm{\s^{-1}}}}
\newcommand{\pMpc}{\ensuremath{\mathrm{\Mpc^{-1}}}}
\newcommand{\s}{\mbox{\ensuremath{\mathrm{~s}}}}
\newcommand{\Mpc}{\mbox{\ensuremath{\mathrm{~Mpc}}}}
\newcommand{\Chandra}{\textit{Chandra}}
\newcommand{\halpha}{\ensuremath{\mathrm{H}\alpha}}
\newcommand{\keV}{\mbox{\ensuremath{\mathrm{~keV}}}}
\newcommand{\cm}{\mbox{\ensuremath{\mathrm{~cm}}}}
\newcommand{\cmsq}{\ensuremath{\mathrm{\cm^2}}}
\newcommand{\ent}{\ensuremath{\mathrm{\keV \cmsq}}}
\newcommand{\kpc}{\mbox{\ensuremath{\mathrm{~kpc}}}}
\newcommand{\kna}{\ensuremath{K_0}}
\newcommand{\khun}{\ensuremath{K_{100}}}
\newcommand{\eg}{{\textit{e.g.}}}
\newcommand{\ie}{{\textit{i.e.}}}
\newcommand{\LCDM}{$\Lambda$CDM~}
\newcommand{\Hn}{\ensuremath{H_0}}
\newcommand{\OM}{\ensuremath{\Omega_{\mathrm{M}}}}
\newcommand{\OL}{\ensuremath{\Omega_{\mathrm{\Lambda}}}}
\newcommand{\clnum}{\ensuremath{222}}
\newcommand{\accept}{\textit{ACCEPT}}
\newcommand{\kthr}{\ensuremath{K_{\mathrm{thresh}}}}
\newcommand{\fha}{\ensuremath{\kna = 13.9 \pm 4.9 \ent}}
\newcommand{\nfha}{\ensuremath{\kna = 130 \pm 55 \ent}}
\newcommand{\frad}{\ensuremath{\kna = 18.3 \pm 7.7 \ent}}
\newcommand{\nfrad}{\ensuremath{\kna = 112 \pm 45 \ent}}

\documentclass{emulateapj}
\usepackage{apjfonts,graphicx,here,lscape}
\begin{document}
\title{An Entropy Threshold for Strong \halpha\ and Radio Emission\\in the Cores of Galaxy Clusters}
\author{
  Kenneth. W. Cavagnolo,
  Megan Donahue,
  G. Mark Voit,
  and Ming Sun}
\affil{Michigan State University, Department of Physics and Astronomy, East Lansing, MI 48824; cavagnolo@pa.msu.edu}
\shorttitle{Entropy Threshold for Feedback}
\shortauthors{K. W. Cavagnolo et al.}

\begin{abstract}
  Our \Chandra\ X-ray Observatory archival study of intracluster
  entropy in a sample of \clnum\ galaxy clusters shows that
  \halpha\ and radio emission from the brightest cluster galaxy are
  much more pronounced when the cluster's core gas entropy is $\la 30
  \ent$. The prevalence of \halpha\ emission below this threshold
  indicates that it marks a dichotomy between clusters that can harbor
  multiphase gas and star formation in their cores and those that
  cannot. The fact that strong central radio emission also appears
  below this boundary suggests that AGN feedback turns on when the
  intracluster medium starts to condense, strengthening the case for
  AGN feedback as the mechanism that limits star formation in the
  Universe's most luminous galaxies.
\end{abstract}

\keywords{conduction -- cooling flows -- galaxies: active -- galaxies:
  clusters: general -- X-rays: galaxies: clusters -- galaxies:
  evolution}

\section{Introduction}
\label{sec:intro}

In recent years the ``cooling flow problem'' has been the focus of
intense scrutiny as the solutions have broad impact on our theories of
galaxy formation (see \citealt{cfreview} for a review). Current models
predict that the most massive galaxies in the Universe -- brightest
cluster galaxies (BCGs) -- should be bluer and more massive than
observations find, unless AGN feedback intervenes to stop late-time
star formation \citep{bower06, croton06, saro06}. X-ray observations
of galaxy clusters have given this hypothesis considerable
traction. From the properties of X-ray cavities in the intracluster
medium (ICM), \cite{birzan04} concluded that AGN feedback provides the
necessary energy to retard cooling in the cores of clusters (see
\citealt{mcnamrev} for a review). This result suggests that, under the
right conditions, AGN are capable of quenching star formation by
heating the surrounding ICM.

If AGN feedback is indeed responsible for regulating star formation in
cluster cores, then the radio and star-forming properties of galaxy
clusters should be related to the distribution of ICM specific
entropy\footnote{In this paper we quantify entropy in terms of the
  adiabatic constant $K = kTn_e^{-2/3}$.}. In previous observational
work (see \citealt{radioquiet, d06, accept}), we have focused on ICM
entropy as a means for understanding the cooling and heating processes
in clusters because it is a more fundamental property of the ICM than
temperature or density alone \citep{voitbryan,voitreview}. ICM
temperature mainly reflects the depth and shape of the dark matter
potential well, while entropy depends more directly on the history of
heating and cooling within the cluster and determines the density
distribution of gas within that potential.

We have therefore undertaken a large \Chandra\ archival project to
study how the entropy structure of clusters correlates with other
cluster properties. \cite{accept} presents the radial entropy profiles
we have measured for a sample of \clnum\ clusters taken from the
\Chandra\ Data Archive. We have named this project the Archive of
Chandra Cluster Entropy Profile Tables, or \accept\ for short. To
characterize the ICM entropy distributions of the clusters, we fit the
equation $K(r) = K_0 +K_{100}(r/100 \kpc)^{\alpha}$ to each entropy
profile. In this equation, \khun\ is the normalization of the
power-law component at 100 kpc and we refer to \kna\ as the central
entropy. Bear in mind, however, that \kna\ is not necessarily the
minimum core entropy or the entropy at $r=0$, nor is it the gas
entropy which would be measured immediately around the AGN or in a BCG
X-ray coronae. Instead, \kna\ represents the typical excess of core
entropy above the best fitting power-law found at larger
radii. \cite{accept} shows that \kna\ is non-zero for almost all
clusters in our sample.

In this letter we present the results of exploring the relationship
between the expected by-products of cooling, \eg\ \halpha\ emission,
star formation, and AGN activity, and the \kna\ values of clusters in
our survey. To determine the activity level of feedback in cluster
cores, we selected two readily available observables: \halpha\ and
radio emission. We have found that there is a critical entropy level
below which \halpha\ and radio emission are often present, while above
this threshold these emission sources are much fainter and in most
cases undetected. Our results suggest that the formation of thermal
instabilities in the ICM and initiation of processes such as star
formation and AGN activity are closely connected to core entropy, and
we suspect that the sharp entropy threshold we have found arises from
thermal conduction (\citealt{conduction}, in press).

This letter proceeds in the following manner: In \S\ref{sec:data} we
cover the basics of our data analysis. The
entropy-\halpha\ relationship is discussed in \S\ref{sec:sf}, while
the entropy-radio relationship is discussed in \S\ref{sec:agn}. A
brief summary is provided in \S\ref{sec:diss}. For this letter we
have assumed a flat \LCDM\ Universe with cosmogony $\OM=0.3$,
$\OL=0.7$, and $\Hn=70\km\ps\pMpc$. All uncertainties are 90\%
confidence.

\section{Data Analysis}
\label{sec:data}

This section briefly describes our data reduction and methods for
producing entropy profiles. More thorough explanations are given in
\cite{d06}, \cite{accept}, and \cite{xrayband}.

\subsection{X-ray}
\label{sec:xray}

X-ray data was taken from publicly available observations in the
\Chandra\ Data Archive. Following standard \Ciao\ reduction
techniques\footnote{http://cxc.harvard.edu/ciao/guides/}, data was
reprocessed using \Ciao\ 3.4.1 and \Caldb\ 3.4.0, resulting in point
source and flare clean events files at level-2. Entropy profiles were
derived from the radial ICM temperature and electron density profiles.

Radial temperature profiles were created by dividing each cluster into
concentric annuli with the requirement of at least three annuli
containing a minimum of 2500 counts each. Source spectra were
extracted from these annuli, while corresponding background spectra
were extracted from blank-sky backgrounds tailored to match each
observation. Each blank-sky background was corrected to account for
variation of the hard-particle background, while spatial variation of
the soft-galactic background was accounted for through addition of a
fixed background component during spectral fitting. Weighted responses
which account for spatial variations of the CCD calibration were also
created for each observation. Spectra were then fit over the energy
range 0.7-7.0 keV in \xspec\ 11.3.2ag \citep{xspec} using a
single-component absorbed thermal model.

Radial electron density profiles were created using surface brightness
profiles and spectroscopic information. Exposure-corrected,
background-subtracted, point-source-clean surface brightness profiles
were extracted from $5\arcsec$ concentric annular bins over the energy
range 0.7-2.0 keV. In conjunction with the spectroscopic normalization
and 0.7-2.0 keV count rate, surface brightness was converted to
electron density using the deprojection technique of
\cite{kriss83}. Errors were estimated using 5000 Monte Carlo
realizations of the surface brightness profile.

A radial entropy profile for each cluster was then produced from the
temperature and electron density profiles. The entropy profiles were
fit with a simple model which is a power-law at large radii and
approaches a constant value, \kna, at small radii (see
\S\ref{sec:intro} for the equation). We define central entropy as
\kna\ from the best-fit model.

\subsection{\halpha}
\label{sec:ha}

One goal of our project was to determine if ICM entropy is connected
to processes like star formation. Here we do not directly measure star
formation but instead use \halpha, which is usually a strong indicator
of ongoing star formation in galaxies \citep{kennicuttrelation}. It is
possible that some of the \halpha\ emission from BCGs is not produced
by star formation \citep{begelman90, sparks04, rusz08,
 ferland08}. Nevertheless, \halpha\ emission unambiguously indicates
the presence of $\sim 10^4$ K gas in the cluster core and therefore
the presence of a multiphase intracluster medium that could
potentially form stars.

Our \halpha\ values have been gathered from several sources, most
notably \cite{crawford99}. Additional sources of data are M. Donahue's
observations taken at Las Campanas and Palomar (reported in
\citealt{accept}), \cite{heckman89}, \cite{dsg92}, \cite{lawrence96},
\cite{1996AJ....112.1390V}, \cite{white97},
\cite{2005MNRAS.363..216C}, and \cite{quillen08}. We have recalculated
the \halpha\ luminosities from these sources using our assumed
\LCDM\ cosmological model.  However, the observations were made with a
variety of apertures and in many cases may not reflect the full
\halpha\ luminosity of the BCG. The exact levels of \lha\ are not
important for the purposes of this letter and we use the \lha\ values
here as a binary indicator of multiphase gas: either \halpha\ emission
and cool gas are present or they are not.

\subsection{Radio}
\label{sec:radio}

Another goal of this work was to explore the relationship between ICM
entropy and AGN activity. It has long been known that BCGs are more
likely to host radio-loud AGN than other cluster galaxies
\citep{burns81, valentijn83, burns90}. Thus, we chose to interpret
radio emission from the BCG of each \accept\ cluster as a sign of AGN
activity.

To make the radio measurements, we have taken advantage of the nearly
all-sky flux-limited coverage of the NRAO VLA Sky Survey (NVSS,
\citealt{nvss}) and Sydney University Molonglo Sky Survey (SUMSS,
\citealt{sumss1, sumss2}). NVSS is a continuum survey at 1.4 GHz of
the entire sky north of $\delta = -40\degr$, while SUMSS is a
continuum survey at 843 MHz of the entire sky south of $\delta =
-30\degr$. The completeness limit of NVSS is $\approx 2.5$ mJy and for
SUMSS it is $\approx 10$ mJy when $\delta > -50\degr$ or $\approx 6$
mJy when $\delta \leq -50\degr$. The NVSS positional uncertainty for
both right ascension and declination is $\la 1''$ for sources brighter
than 15 mJy, and $\approx 7''$ at the survey detection limit
\citep{nvss}. At $z=0.2$, these uncertainties represent distances on
the sky of $\sim3-20$ kpc. For SUMSS, the positional uncertainty is
$\la 2''$ for sources brighter than 20 mJy, and is always less than
$10''$ \citep{sumss1,sumss2}. The distance at $z=0.2$ associated with
these uncertainties is $\sim6-30$ kpc. We calculate the radio power
for each radio source using the standard relation $\nu L_{\nu} = 4 \pi
D_L^2 S_{\nu} f_0$ where $S_{\nu}$ is the 1.4 GHz or 843 MHz flux from
NVSS or SUMSS, $D_L$ is the luminosity distance, and $f_0$ is the
central beam frequency of the observations. Our calculated radio
powers are simply an approximation of the bolometric radio luminosity.

Radio sources were found using two methods. The first method was to
search for sources within a fixed angular distance of $20\arcsec$
around the cluster X-ray peak. The probability of randomly finding a
radio source within an aperture of $20\arcsec$ is exceedingly low ($<
0.004$ for NVSS). Thus, in \clnum\ total field searches, we expect to
find no more than one spurious source. The second method involved
searching for sources within 20 projected kpc of the cluster X-ray
peak. At $z \approx 0.051$, $1\arcsec$ equals 1 kpc, thus for clusters
at $z \ga 0.05$, the 20 kpc aperture is smaller than the $20\arcsec$
aperture, and the likelihood of finding a spurious source gets
smaller. Both methods produce nearly identical lists of radio sources
with the differences arising from the very large, extended lobes of
low-redshift radio sources such as Hydra A.

To make a spatial and morphological assessment of the radio emission's
origins, \ie\ determining if the radio emission is associated with the
BCG, high angular resolution is necessary. However, NVSS and SUMSS are
low-resolution surveys with FWHM of $\approx 45\arcsec$. We therefore
cannot distinguish between ghost cavities/relics, extended lobes,
point sources, re-accelerated regions, or if the emission is coming
from a galaxy very near the BCG. We have handled this complication by
visually inspecting each radio source in relation to the optical
(using DSS I/II\footnote{http://archive.stsci.edu/dss/}) and infrared
(using 2MASS\footnote{http://www.ipac.caltech.edu/2mass/}) emission of
the BCG. We have used this method to establish that the radio emission
is most likely coming from the BCG. When available, high resolution
data from VLA FIRST\footnote{http://sundog.stsci.edu} was added to the
visual inspection. VLA FIRST is a 10,000$\square^\circ$
high-resolution ($5\arcsec$) survey at 20 cm of the North and South
Galactic caps \citep{first}. FIRST is also more sensitive than either
NVSS or SUMSS with a detection threshold of 1 mJy.

\section{\halpha\ Emission and Central Entropy}
\label{sec:sf}

\begin{figure}
  \begin{center}
    \includegraphics*[width=\columnwidth, trim=28mm 7mm 40mm 17mm, clip]{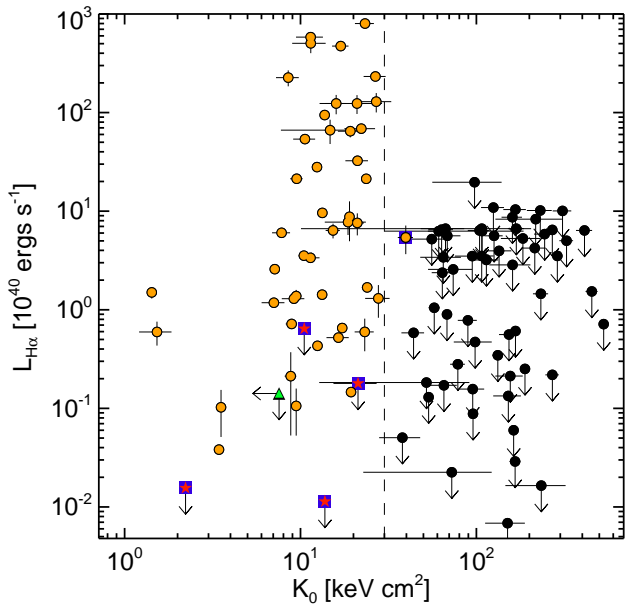}
    \caption{Central entropy vs. \halpha\ luminosity. Orange circles
      represent \halpha\ detections, black circles are non-detection
      upper limits, and blue squares with inset red stars or orange
      circles are peculiar clusters which do not adhere to the
      observed trend (see text). A2151 is plotted using the 2$\sigma$
      upper-limit of the best-fit \kna\ and is denoted by a green
      triangle. The vertical dashed line marks $\kna = 30 \ent$. Note
      the presence of a sharp \halpha\ detection dichotomy beginning
      at $\kna \la 30 \ent$.}
    \label{fig:ha}
  \end{center}
\end{figure}

Of the \clnum\ clusters in \accept, we located \halpha\ observations
from the literature for 110 clusters. Of those 110, \halpha\ was
detected in 46, while the remaining 64 have upper limits. The mean
central entropy for clusters with detections is \fha, and for clusters
with only upper-limits \nfha.

In Figure \ref{fig:ha} central entropy is plotted versus
\halpha\ luminosity. One can immediately see the dichotomy between
clusters with and without \halpha\ emission. If a cluster has a
central entropy $\la 30 \ent$ then \halpha\ emission is usually
``on'', while above this threshold the emission is predominantly
``off''. For brevity we refer to this threshold as
\kthr\ hereafter. The cluster above \kthr\ which has \halpha\ emission
(blue square with inset orange circle) is Zwicky 2701 ($\kna = 39.7
\pm 3.9 \ent$). There are also clusters below \kthr\ without
\halpha\ emission (blue squares with red stars): A2029, A2107, EXO
0422-086, and RBS 533. A2151 also lies below \kthr\ and has no
detected \halpha\ emission, but the best-fit \kna\ for A2151 is
statistically consistent with zero and this cluster is plotted using
the 2$\sigma$ upper-limit of \kna\ (green triangle in
Fig. \ref{fig:ha}).  These five clusters are clearly exceptions to the
much larger trend. The mean and dispersion of the redshifts for
clusters with and without \halpha\ are not significantly different,
$z = 0.124 \pm 0.106$ and $z = 0.132 \pm 0.084$ respectively, and
applying a redshift cut (\ie\ $z = 0-0.15$ or $z = 0.15-0.3$) does not
change the \kna-\halpha\ dichotomy. Most important to note is that
changes in the \halpha\ luminosities because of aperture effects will
move points up or down in Figure \ref{fig:ha}, while mobility along
the \kna\ axis is minimal. Qualitatively, the correlation between low
central entropy and presence of \halpha\ emission is very robust.

The clusters with \halpha\ detections are typically between $10-30
\ent$, have short central cooling times ($<$ 1 Gyr), and under older
nomenclature would be classified as ``cooling flow'' clusters.  It has
long been known that star formation and associated \halpha\ nebulosity
appear only in cluster cores with cooling times less than a Hubble
time \citep{hu85, johnstone87, mcnamara89, voit97,cardiel98}. However,
our results suggest that the central cooling time must be at least a
factor of 10 smaller than a Hubble time for these manifestations of
cooling and star formation to appear.  It is also very interesting
that the characteristic entropy threshold for strong \halpha\ emission
is so sharp. \cite{conduction} have recently proposed electron thermal
conduction may be responsible for setting this threshold. This
hypothesis has received further support from the theoretical work of
\cite{2008arXiv0804.3823G} showing that thermal conduction can
stabilize non-cool core clusters against the formation of thermal
instabilities, and that AGN feedback may be required to limit star
formation when conduction is insufficient.

\section{Radio Sources and Central Entropy}
\label{sec:agn}

\begin{figure}
  \begin{center}
    \includegraphics*[width=\columnwidth, trim=28mm 7mm 40mm 17mm, clip]{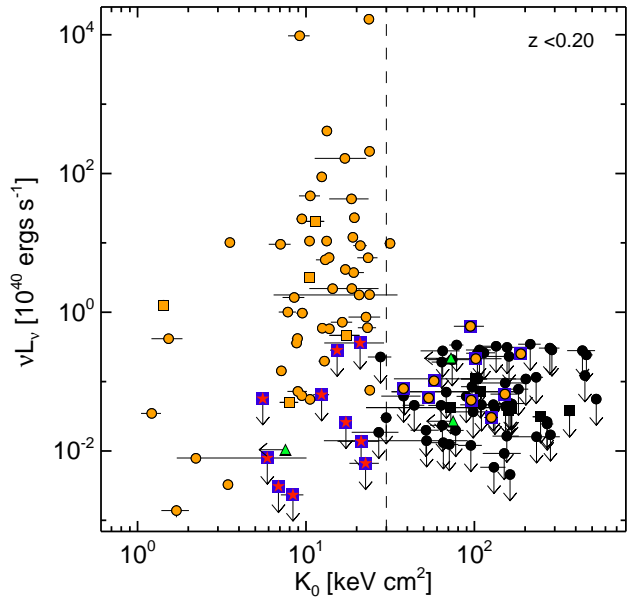}
    \caption{BCG radio power vs. \kna\ for clusters with $z <
      0.2$. Orange symbols represent radio detections and black
      symbols are non-detection upper-limits. Circles are for NVSS
      observations and squares are for SUMSS observations. The blue
      squares with inset red stars or orange circles are peculiar
      clusters which do not adhere to the observed trend (see text).
      Green triangles denote clusters plotted using the 2$\sigma$
      upper-limit of the best-fit \kna. The vertical dashed line marks
      $\kna = 30 \ent$. The radio sources show the same trend as
      \halpha: bright radio emission is preferentially ``on'' for
      $\kna \la 30 \ent$.}
    \label{fig:radzcut}
  \end{center}
\end{figure}

Of the \clnum\ clusters in \accept, 100 have radio-source detections
with a mean $\kna$ of $23.3 \pm 9.4 \ent$, while the other 122
clusters with only upper limits have a mean $\kna$ of $134 \pm 52
\ent$. NVSS and SUMSS are low resolution surveys with FWHM at $\approx
45\arcsec$ which at $z = 0.2$ is $\approx 150\kpc$. This scale is
larger than the size of a typical cluster cooling region and makes it
difficult to determine absolutely that the radio emission is
associated with the BCG. We therefore focus only on clusters at $z <
0.2$. After the redshift cut, 135 clusters remain -- 64 with radio
detections (mean \frad) and 71 without (mean \nfrad).

In Figure \ref{fig:radzcut} we have plotted radio power versus \kna.
The obvious dichotomy seen in the \halpha\ measures and characterized
by \kthr, is also present in the radio. Clusters with $\nu L_{\nu}
\gtrsim 10^{40} \ergps$ generally have $\kna \la \kthr$. This trend
was first evident in \citet{radioquiet} and suggests that AGN activity
in BCGs, while not exclusively limited to clusters with low core
entropy, is much more likely to be found in clusters which have a core
entropy less than \kthr. That star formation and AGN activity are
subject to the same entropy threshold suggests the mechanism which
promotes or initiates one is also involved in the activation of the
other. If the entropy of the hot gas in the vicinity of the AGN is
correlated with \kna, then the lack of correlation between radio power
and \kna\ below the $30 \ent$ threshold suggests that cold-mode
accretion \citep{pizzolato05, hardcastle07} may be the dominant method
of fueling AGN in BCGs.

We have again highlighted exceptions to the general trend seen in
Figure \ref{fig:radzcut}: clusters below \kthr\ without a radio source
(blue squares with inset red stars) and clusters above \kthr\ with a
radio source (blue squares with inset orange circles). The peculiar
clusters below \kthr\ are A133, A539, A1204, A2107, A2556, AWM7, ESO
5520200, MKW4, MS J0440.5+0204, and MS J1157.3+5531. The peculiar
clusters above \kthr\ are 2PIGG J0011.5-2850, A193, A586, A2063,
A2147, A2244, A3558, A4038, and RBS 461. In addition, there are three
clusters, A2151, AS405, MS 0116.3-0115, which have best-fit
\kna\ statistically consistent with zero and are plotted in
Fig. \ref{fig:radzcut} using the 2$\sigma$ upper-limit of \kna\ (green
triangles). All three of these clusters have detected radio sources.

Finding a few clusters in our sample without radio sources where we
expect to find them is not surprising given that AGN feedback could be
episodic. However, the clusters above \kthr\ with a central radio
source are interesting, and may be special cases of BCGs with embedded
coronae. \cite{coronae} extensively studied coronae and found they are
like ``mini-cooling cores'' with low temperatures and high
densities. Coronae are a low-entropy environment isolated from the
high-entropy ICM and may provide the conditions necessary for gas
cooling to proceed. And indeed, 2PIGG 0011, A193, A2151, A2244, A3558,
A4038, and RBS 461 show indications that a very compact ($r \la
5\kpc$) X-ray source is associated with the BCG \citep{accept}.

\section{Summary}
\label{sec:diss}

We have presented a comparison of ICM central entropy values and
measures of BCG \halpha\ and radio emission for a \Chandra\ archival
sample of galaxy clusters. We find that below a characteristic central
entropy threshold of $\kna \approx 30 \ent$, \halpha\ and bright radio
emission are more likely to be detected, while above this threshold
\halpha\ is not detected and radio emission, if detected at all, is
significantly fainter. The mean \kna\ for clusters with and without
\halpha\ detections are \fha\ and \nfha, respectively. For clusters at
$z < 0.2$ with BCG radio emission the mean \frad, while for BCGs with
only upper limits, the mean \nfrad. While other mechanisms can produce
\halpha\ or radio emission besides star formation and AGN, if one
assumes the \halpha\ and radio emission are coming from these two
feedback sources, then our results suggest the development of
multiphase gas in cluster cores (which can fuel both star formation
and AGN) is strongly coupled to ICM entropy.\\\\

We were supported in this work through CXO grants AR-6016X, AR-4017A,
and NASA LTSA program NNG-05GD82G. The CXC is operated by the SAO for
and on behalf of NASA under contract NAS8-03060.

\end{document}